\documentclass[english,aps,preprint]{revtex4}
\usepackage[T1]{fontenc}
\usepackage[latin9]{inputenc}
\usepackage{array}
\usepackage{float}
\usepackage{amsmath}
\usepackage{graphicx}
\usepackage{amssymb}

\makeatletter

\providecommand{\tabularnewline}{\\}

\@ifundefined{textcolor}{}
{%
 \definecolor{BLACK}{gray}{0}
 \definecolor{WHITE}{gray}{1}
 \definecolor{RED}{rgb}{1,0,0}
 \definecolor{GREEN}{rgb}{0,1,0}
 \definecolor{BLUE}{rgb}{0,0,1}
 \definecolor{CYAN}{cmyk}{1,0,0,0}
 \definecolor{MAGENTA}{cmyk}{0,1,0,0}
 \definecolor{YELLOW}{cmyk}{0,0,1,0}
 }

\@ifundefined{definecolor}
 {\@ifundefined{definecolor}
 {\@ifundefined{definecolor}
 {\@ifundefined{definecolor}
 {\@ifundefined{definecolor}
 {\@ifundefined{definecolor}
 {\@ifundefined{definecolor}
 {\@ifundefined{definecolor}
 {\@ifundefined{definecolor}
 {\@ifundefined{definecolor}
 {\@ifundefined{definecolor}
 {\usepackage{color}}{}
}{}
}{}
}{}
}{}
}{}
}{}
}{}
}{}
}{}
}{}
\makeatother

\makeatother

\usepackage{babel}

\makeatother

\usepackage{babel}

\makeatother

\usepackage{babel}

\makeatother

\usepackage{babel}

\makeatother

\usepackage{babel}

\makeatother

\usepackage{babel}

\makeatother

\usepackage{babel}

\makeatother

\usepackage{babel}

\makeatother

\usepackage{babel}

\makeatother

\usepackage{babel}

\makeatother

\usepackage{babel}

\makeatother

\usepackage{babel}

\begin{document}

\preprint{Manuscript}

\title{Layer and size dependence of thermal conductivity in multilayer graphene
nanoribbons }

\author{Hai-Yuan Cao$^{1}$, Zhi-Xin Guo$^{2}$, Hongjun Xiang$^{1}$ and
Xin-Gao Gong$^{1}$ }

\affiliation{$^{1}$Key Laboratory for Computational Physical Sciences(MOE) and
Surface Physics Laboratory, Fudan Univeristy, Shanghai 200433, China\\
 $^{2}$Department of Physics, Xiangtan University, Xiangtan 411105,
China}
\begin{abstract}
Using non-equilibrium molecular dynamics method(NEMD), we have found
that the thermal conductivity of multilayer graphene nanoribbons monotonously
decreases with the increase of the number of layers, such behavior
can be attributed to the phonon resonance effect of out-of-plane phonon
modes. The reduction of thermal conductivity is found to be proportional
to the layer size, which is caused by the increase of phonon resonance.
Our results are in agreement with recent experiment on dimensional
evolution of thermal conductivity in few layer graphene.


\end{abstract}
\maketitle
The fresh comings of carbon family, the two-dimensional (2D) graphene
and quasi-one-dimensional (1D) graphene nanoribbon (GNR), have attracted
strong interest due to their fundamental physical properties and potential
applications in nanoelectronic devices\cite{1,2,3,4}. Experimental
developments have enabled the growth of the high quality multilayer
graphene\cite{5} and the control of GNR edge geometries\cite{6}.
As we know, the thermal property is a crucial aspect that determines
the application of a material in the nanoelectronics\cite{7}. In
recent years, the thermal properties of monolayer graphene and GNRs
have aroused much attention \cite{8,9,10,11,12,13,14,15,16,17}. Very
recently, the thermal conductivity of few layer graphene has been
also studied\textbf{ }experimentally and theoretically\cite{5}. However,
the thermal conductivity of multilayer GNRs have not been well investigated
so far. Its clarification is becoming desirable due to the forthcoming
application of multilayer GNRs in nanoelectronics\cite{18,19,20}.

On the other hand, the mechanism of thermal conduction on the nanoscale
is currently a controversial issue\cite{21,7}. Investigating the
dimensional evolution of thermal conductivity could provide a new
insight to clarify the fundamental mechanism on nanoscale. The thermal
conductivity evolution from 2D to 3D had been investigated experimentally\cite{5},
while the evolution from 1D to higher dimensions is still not well
studied. The multilayer GNRs are ideal systems for such investigation,
whose dimensional evolution can be easily realized through the layer
number and size variation.

In this letter, we employ the non-equilibrium molecular dynamics
(NEMD) method to investigate layer and size effect on thermal
conductivity of multilayer GNRs with different edge shapes, such as
the armchair multilayer GNRs and the zigzag multilayer GNRs. The
number of layers varies from 1 to 4 and the width of layers varies
from 1 to 10 nm. We find that both the number of layers and size
have strong influence on thermal conductivity of multilayer GNRs.
The intrinsic mechanism of thermal conductivity variation is further
explored through\textbf{ }a phonon resonance model.

The structure of multilayer GNRs are constructed based on the theoretical
prediction of bilayer GNRs which has a small deviation from Bernal
stacking\cite{22}. In the NEMD simulation, Tersoff potential\cite{23}
is utilized to describe the in-plane C-C bonding interactions and
Lennard-Jones potential is used to describe the intra-plane van der
Waals interactions\cite{24}. We use the velocity Verlet method to
integrate equations of motion with a fixed time step of 1 fs\cite{17,26,27,25}.
On each layer of multilayer GNR\textbf{,} fixed boundary condition
is implemented\textbf{ }with the atoms at the left and right ends\textbf{
}fixed at their equilibrium positions\cite{16,17}. Next to the boundaries,
the adjacent two cells of atoms are coupled to $Nos\acute{e}-Hoover$
thermostats with temperatures 310 K and 290 K, respectively. The thermal
conductivity for layer $i$ can be calculated directly from the well
known Fourier law $\kappa_{i}=J_{i}d/(\Delta Twh)$, where\textbf{
$\Delta T=$}20 K is the temperature difference between two thermostats,
$J_{i}$ is the heat flux from the heat bath to the system , which
can be obtained via calculating the power of heat baths\cite{25},
$d$ is the length, $w$ is the width, $h$ (\textbf{ }0.144 nm )\cite{17}
is the thickness of a monolayer GNR. All the calculated thermal conductivities
are obtained by averaging about 3 ns after 2 ns to establish a stable
temperature gradient along the length direction. Thus the thermal
conductivity of multilayer GNR can be defined as $\kappa=\sum\kappa_{i}/n$
with $n$ being the\textbf{ }number of layers. In addition, all the
GNR structures have been optimized before NEMD simulation. We also
define a multilayer GNR with N carbon-chains in width to be represented
as $\textit{N}$-armchair multilayer GNR or $\textit{N}$-zigzag multilayer
GNR, depending on the specific edge shapes\cite{17}.

We first investigate the number of layers dependence of thermal conductivity
of 20-armchair multilayer GNR and 10-zigzag multilayer GNR. Similar
to that of monolayer GNRs, the 20-armchair multilayer GNRs have much
lower thermal conductivity than 10-zigzag multilayer GNRs (see Fig.
1), implying a universal edge-shape dependence of thermal conductivity
in the GNR family. Moreover, with the number of layers increasing,
thermal conductivity of both the armchair multilayer GNR and zigzag
multilayer GNR monotonously decreases.\textbf{ }When the number of
layers gets to 4, the thermal conductivity is reduced to 123 Wm$^{-1}$K$^{-1}$
and 308 Wm$^{-1}$K$^{-1}$ for 20-armchair multilayer GNR and 10-zigzag
multilayer GNR, respectively\textbf{.} Comparing with that of monolayer
20-armchair GNR (195 Wm$^{-1}$K$^{-1}$) and 10-zigzag GNR (495 Wm$^{-1}$K$^{-1}$),
the reduction of thermal conductivity of armchair multilayer GNR and
zigzag multilayer GNR gets\textbf{ }to 40\%, showing an obvious dependence
on number of layers. This indicates that the crossplane coupling is
enhanced with number of layers increasing, and it plays an important
role in the evolution of thermal conductivity from monolayer GNR to
multilayer GNR\textbf{. }This is in agreement with recent experiment
in few layer graphene, where a 67\% reduction of thermal conductivity
is observed as the number of atomic plane increasing from 1 to 4\cite{5}.

These results can be further understood by the coupling mechanism\cite{28},
there exists a competitive mechanism on thermal conductivity in a
coupling system: the phonon-resonance effect that decreasing thermal
conductivity and phonon-band-up-shift effect that increasing thermal
conductivity \cite{28}. The strength of phonon resonance can be described
by the resonance angle\textbf{ $\Psi$, }which is determined by atomic
mass and spring constant of two coupled systems. When $\Psi$ is small($\Psi<\dfrac{\pi}{24}$),
the variation of thermal conductivity is dominated by the phonon-band-up-shift
effect; when $\Psi$ comes to large($\Psi>\dfrac{\pi}{12}$), the
thermal conductivity is dominated by the phonon resonance effect.
The thermal conductivity reduction of multilayer GNRs with the increase
of number of layers can be accounted for this mechanism. For the multilayer
GNRs, the atomic mass and spring constant of the coupling layers are
totally equivalent, thus\textbf{ $\Psi=\dfrac{\pi}{4}$ }and the phonon-resonance
effect plays a dominant role on thermal conductivity variation. Therefore,
the thermal conductivity of multilayer GNRs monotonously decreases
with the number of layers increasing which induces more and more intensive
phonon resonance.

In order to identify the phonon-resonance effect in multilayer GNRs,
we freeze the out-of-plane atomic vibration of multilayer GNRs and
re-calculate their thermal conductivity. For simplicity, we consider
the bilayer GNRs. Here we only present the calculated thermal conductivity
values for the zigzag bilayer GNR, the results for armchair bilayer
GNR are qualitatively similar. Two cases are considered: out-of-plane
vibration of the top layer is frozen(constraint 1) and out-of-plane
vibration of both layers is frozen(constraint 2).

As shown in Table 1, freezing out-of-plane atomic vibration would
considerably change the layer's thermal conductivity. If only top
layer's out-of-plane vibration is frozen, thermal conductivity of
top layer would increase by 40\% (from 334 Wm$^{-1}$K$^{-1}$ to
467 Wm$^{-1}$K$^{-1}$) while thermal conductivity of bottom layer
is nearly unaffected by the artificial constraint. If the out-of-plane
vibration is frozen in both layers, thermal conductivity of bilayer
zigzag GNR almost equals to that of monolayer zigzag GNR. This indicates
that, in the multilayer GNRs, the phonon resonance of stacking layers
is mainly from the coupling between the crossplane out-of-plane ZO'
phonon modes\cite{29} and the out-of-plane phonon modes that propagate
in the basal plane.

For the purpose of investigating finite size effect, we calculate
thermal conductivity of both monolayer and bilayer GNRs with various
width of GNRs. Figure 2 shows the obtained thermal conductivity of
monolayer and bilayer GNRs, whose width ranges from 1 nm to 10 nm.
As one can see, the thermal conductivity of zigzag bilayer GNR increases
firstly and then turns to decrease with the width increasing, while
the armchair bilayer GNR's thermal conductivity monotonously increases
with the width increasing. The width dependence of thermal conductivity
of bilayer GNRs is very similar to that of monolayer GNRs (Fig. 2).
This phenomena can be attributed to the competition between edge-localized
phonon effect and the umklapp phonon scattering effect with the width
of ribbon variation\cite{17}.

In addition, the difference of thermal conductivity between monolayer
and bilayer GNRs also exhibits an obvious layer-size dependence. For
the armchair GNRs (zigzag GNRs), the reduction of thermal conductivity
monotonously increases from 31\% (30\%) to 35\% (32\%) with the width
increasing from 1 nm to 10 nm.\textbf{ }This means the difference
of thermal conductivity between monolayer GNR and multilayer GNR is
proportional to the layer size, because the phonon resonance strength
between different layers is proportional to the number of total phonon
modes which is in turn corresponding to the size of ribbon. Our results
on thermal conducitivity reduction of multilayer GNRs is smaller than
that of few layer graphenes\cite{5}. Above results indicate that
the difference in dimension evolution of thermal conductivity between
multilayer graphene and multilayer GNRs comes from the finite size
effect.

In conclusion, we have investigated the thermal conductivity of multilayer
GNRs using the NEMD method. Comparing to monolayer GNR, the thermal
conductivity of multilayer GNRs has a significant reduction which
is in agreement with available experiments\cite{5}. Based on the
phonon resonance model, we propose that the reduction of thermal conductivity
is attributed to the resonance of out-of-plane phonon modes. Moreover,
the difference between thermal conductivities of GNRs with different
number of layers is found to be proportional to the layer size, which
is directly determined by the number of the total phonon modes. The
present studies suggest that the thermal conductivity of multilayer
GNRs can be manipulated by changing the number and size of layers,
which provides potential applications of multilayer GNR-based materials
in future nanodevices.

This work was partially supported by the Special Funds for Major State
Basic Research , National Science Foundation of China , Ministry of
Education and Shanghai Municipality. The computation was performed
in the Supercomputer Center of Shanghai, the Supercomputer Center
of Fudan University.

\clearpage{}

\clearpage{} %
\begin{table}[H]
 \caption{Thermal conductivity $\kappa$ of bilayer zigzag GNRs with different
constraints. The length and width of the monolayer and bilayer zigzag
GNR is about 10 nm and 3 nm.}

\begin{ruledtabular} \begin{tabular}{cc>{\centering}p{1in}>{\centering}p{1in}}
Type of GNR  & Constraint  & \multicolumn{2}{c}{Thermal Conductivity (Wm$^{-1}$K$^{-1}$)}\tabularnewline
 &  & Top Layer  & Bottom layer \tabularnewline
Bilayer zigzag GNR  & Free  & 323  & 335 \tabularnewline
Bilayer zigzag GNR  & Constraint 1  & 330  & 467\tabularnewline
Bilayer zigzag GNR  & Constraint 2  & 469  & 480 \tabularnewline
Monolayer zigzag GNR  & Free  & 490  & --\tabularnewline
\end{tabular}\end{ruledtabular}

{\scriptsize Constraint 1: Out-of-plane modes of top layer are frozen
$\ $Constraint 2: Out-of-plane modes of both layers are frozen }
\end{table}

\clearpage{} %
\begin{figure}
\begin{centering}
\includegraphics[scale=0.4,angle=-90]{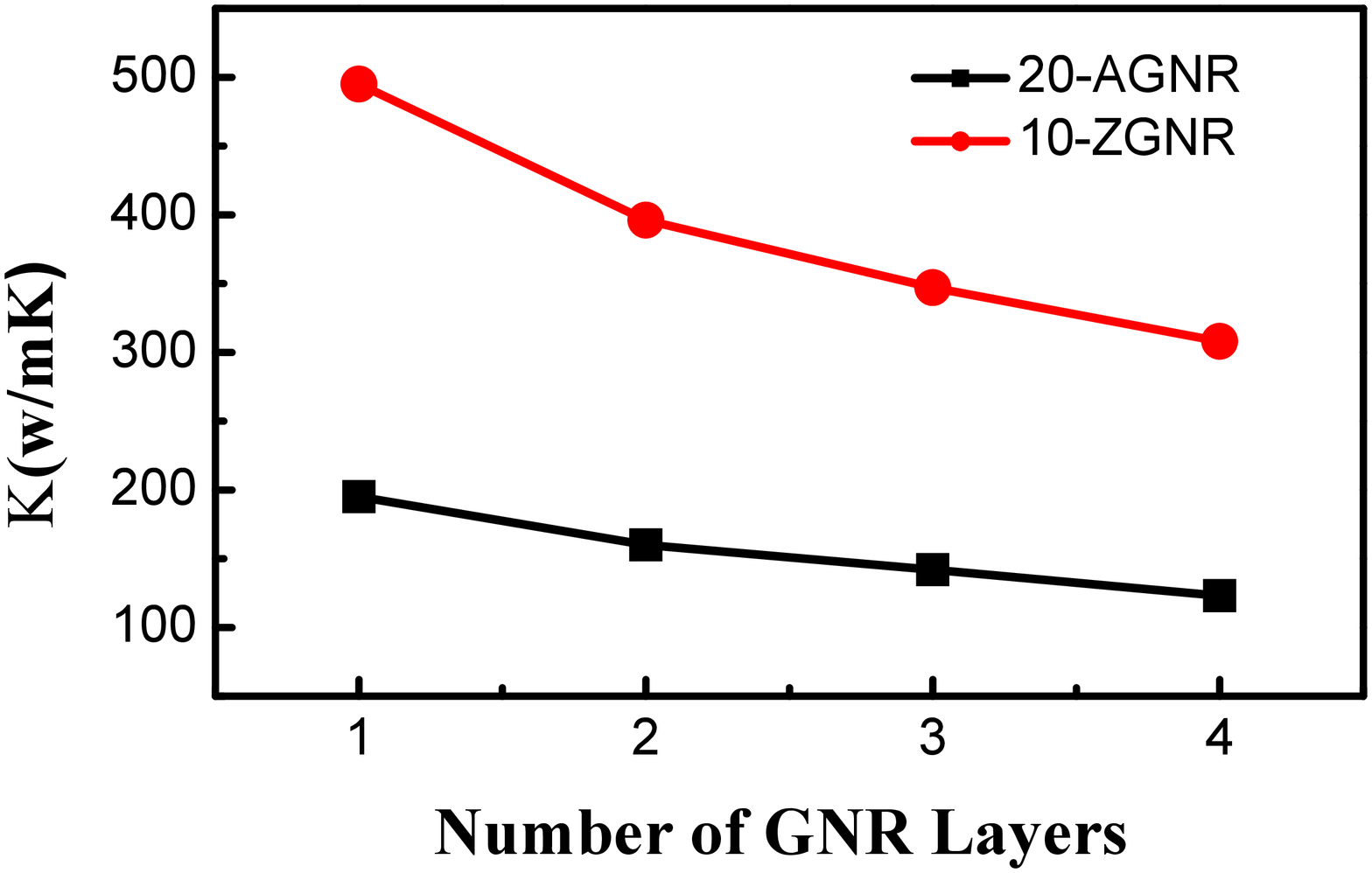}
\end{centering} \caption{Thermal conductivity $\kappa$ as a function
of the number of layers with length and width of 5 nm and 2 nm.
$\kappa$ decreases monotonously with number of GNR layers
increasing, indicating the enhancement of intra-layer phonon
coupling.}
\end{figure}

\clearpage{} %
\begin{figure}
\begin{centering}
\includegraphics[scale=0.4,angle=-90]{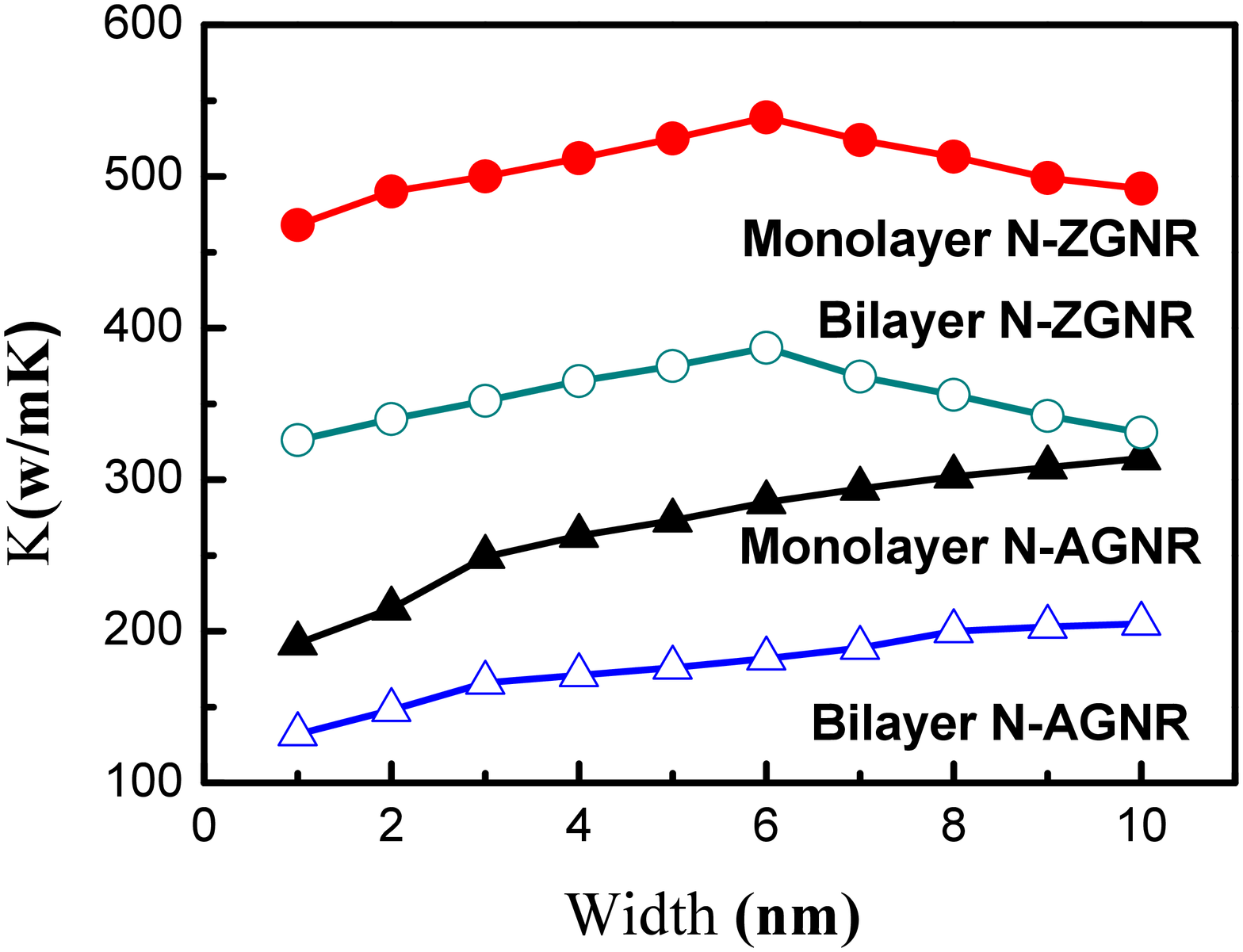}
\end{centering}
\caption{Thermal conductivity $\kappa$ as a function of the width of
monolayer or bilayer armchair GNR and zigzag GNR with fixed length
of 10 nm. The behavior of $\kappa$ for monolayer and bilayer is
similar and the difference of $\kappa$ between monolayer and bilayer
GNRs increases with the width of GNRs increasing. }
\end{figure}

\end{document}